\documentclass[preprint]{aastex}

\slugcomment{Submitted to Astrophysical Journal Letters}
\shorttitle{Poisson hypothesis for flares}
\shortauthors{Wheatland}

\begin{document}

\title{The local Poisson hypothesis for solar flares}

\author{M.S. Wheatland}

\affil{School of Physics, The University of Sydney,
  NSW 2006, Australia}
\email{wheat@physics.usyd.edu.au} 

\begin{abstract}
The question of whether flares occur as a Poisson process has
important consequences for flare physics. Recently Lepreti et al.\ 
presented evidence for local 
departure from Poisson statistics in the Geostationary Operational 
Environmental Satellite (GOES) X-ray flare catalog. Here it is argued that 
this effect arises from a selection effect inherent in the soft X-ray 
observations; namely that the slow decay of enhanced flux following a large 
flare makes detection of subsequent flares less likely. It is also shown 
that the power-law tail of the GOES waiting-time distribution varies with 
the solar cycle. This counts against any intrinsic significance to the 
appearance of a power law, or to the value of its index.
\end{abstract}


\section{Introduction}

\noindent
There has been increased recognition of the importance of flare 
statistics for understanding the mechanisms of energy storage and
release in the solar corona. The distribution of times $\Delta t$ between 
events (``waiting times'') provides information on whether flares occur as 
independent events, and also represents a test for certain flare models.
In particular, the avalanche model for flares \citep{lu&ham91,lu&93} and 
variants thereof, predict that flares are independent and occur with a 
constant rate, and hence obey Poisson statistics. The waiting-time 
distribution (WTD) for a Poisson process with a rate $\lambda$ is an 
exponential,
\begin{equation}
P(\Delta t)=\lambda e^{-\lambda \Delta t}.
\end{equation}   
  
Observational determinations of the flare WTD based on different data sets
have given a variety of results (see e.g.\ Wheatland et al.\ 1998 for a 
brief review). Boffeta et al.\ (1999) showed that the WTD
constructed from the GOES catalog of 20 years of flaring exhibits a
power-law tail. They argued that the appearance of a power law contradicts
the avalanche model prediction of Poisson statistics, and supports instead
an MHD turbulence model of the flaring process. Special significance was
attributed to the value of the power-law index ($\alpha\approx -2.4$) 
derived from the observations. In response, Wheatland (2000) argued that 
over the course of the several solar cycles
included in the GOES data, the flaring rate varies by more than an order of
magnitude, and so flares cannot be assumed to be occurring at a constant
rate. If the flaring process can be represented by a piecewise constant
Poissson process with constant rates $\lambda_i$ for intervals $t_i$, then
the WTD may be described by
\begin{equation}\label{eq:varying_poisson}
P(\Delta t)=\sum_i\varphi_i \lambda_i e^{-\lambda_i\Delta t},
\end{equation}
where $\varphi_i=\lambda_i t_i/\sum_i\lambda_i t_i$ 
is the fraction of events corresponding to 
the rate $\lambda_i$. Rates $\lambda_i$ and intervals $t_i$ were 
estimated for the GOES catalog events using a Bayesian procedure. 
The resulting WTD [equation~(\ref{eq:varying_poisson})] was shown to 
qualitatively
reproduce the observed power-law tail. The GOES catalog involves flares
occurring in all active regions on the Sun. Wheatland (2001) 
extended the analysis to flares in individual active regions, and showed
that equation~(\ref{eq:varying_poisson}) accounts quantitatively for the 
observed WTDs in a number of very flare-productive active regions. 

Recently Lepreti et al.\ (2001) responded by showing that 
a statistical test rejects the hypothesis that the GOES catalog events 
are locally Poisson. They showed that the flare WTD can be fitted by a 
L\'{e}vy function, and argued that this reflects the existence of memory 
in the system and once again supports a turbulence model for the flare 
process.

Although the question of whether flares occur as a Poisson process may 
seem arcane, the consequences for the understanding of the flare process
are substantial.

In this paper the result of Lepreti et al.\ (2001) is re-examined. 
In \S\,2 an alternative explanation for the local departure of the 
GOES flares from 
Poisson statistics is given, namely that it arises from the failure to 
detect flares occurring soon after large flares because of the increase in 
soft X-ray flux associated with the large flare. The nature of soft
X-ray data and the event selection procedure used to compile the GOES 
catalog makes this effect inevitable. In \S\,3 it is shown that the 
power-law tail of the GOES WTD varies with the solar cycle. This result
argues against any particular significance to the value of the power-law
index of the tail of the WTD. Finally in \S\,4 the consequences of these 
results to our understanding of flares are discussed.
 
\section{Local departure from Poisson statistics}

\noindent
Lepreti et al.\ (2001) applied a test for local Poisson behaviour devised 
by Bi et al.\ (1989) to the GOES catalog events. Flares are replaced by
their peak times. Let $X_i$ denote the time interval between the peak time
of event $i$ and that of its nearest neighbour. Also, let $Y_i$ denote the time
interval to the {\em other} neighbour. Under the local Poisson hypothesis, 
$X_i$ and $Y_i$ have probability distributions 
$P(X_i)=2\lambda_i \exp (-2\lambda_i X_i)$ and 
$P(Y_i)=\lambda_i \exp(-\lambda_i Y_i)$. In this case the variable
\begin{equation}
H_i=\frac{X_i}{X_i+\frac{1}{2}Y_i}
\end{equation}
should be uniformly distributed between zero and one, and then the observed
distribution of $H_i$ provides a test of the local Poisson hypothesis.
Lepreti et al.\ (2001) reported a significantly non-uniform distribution of 
$H_i$ for the GOES events. We have repeated the procedure for GOES events 
between 1981-1999 above C1 class (for details of the GOES data, see
Wheatland 2001), and the result is shown in Fig.~1. The solid curve is
the cumulative distribution of $H_i$. A Kolmogorov-Smirnov test rejects 
the hypothesis that the distribution of $H_i$ is uniform, confirming the 
finding of Lepreti et al.\ (2001).

Of course, the departure of the GOES catalog events from Poisson behaviour
does not necessarily mean that solar flares are not locally Poisson. It is 
important to consider how the catalog was constructed, and in particular 
any selection biases that might introduce a departure from Poisson
statistics. Recently Wheatland (2001) identified such an effect,
which was termed {\em obscuration}. The GOES detectors record soft X-ray
light curves for flares, which are characterised by relatively fast rises 
to a peak flux, followed by slow decays in flux due to the cooling of 
hot plasma in the corona produced by the flare \citep{fel96}. 
Large flares produce 
enhancements of more than a factor of 100 in flux, and this increase
may take many hours to decline. During the decay time subsequent flares 
may be missed because of the enhanced background. The selection
procedure used to compile the GOES catalog imposes constraints on the 
detection of flares. The start of a flare is defined by four one-minute
flux values that are monotonically increasing, with the final flux 
being at least $1.4$ times the first. For flares occurring shortly after a 
large flare, this
implies that the second flare must produce an increase in flux at least 
of order $40\%$ of the flux of the first flare to be detected. Even quite 
significant events can be missed in the
wake of a large flare. Since flares follow a power-law peak flux
distribution \citep{hud91}, the majority of flares are small and a large 
number of flares are expected to be missing from the catalog as a result. 
Wheatland (2001) estimate that in the absence of obscuration the
number of flares above C1 class would be higher by $(75\pm 23)\%$.  

The following simulation confirms that obscuration can account for the 
observed departure from Poisson statistics. Using the Bayesian
decomposition of the GOES catalog into rates $\lambda_i$ and intervals 
$t_i$ described in Wheatland (2000), a sequence of flare times
was generated as a piecewise constant Poisson process.
As a check, the Bi et al.\ (1989) test was applied to this sequence of 
times, and local Poisson behaviour was confirmed (the cumulative
distribution of $H_i$ is shown by the dotted line in Fig.~1).
Next, a decay time was assigned to each synthetic flare, based on random 
selection from the decay times observed for the real GOES events. To mimic
obscuration, any events falling within the decay time of another event were 
excluded from the sequence. The Bi et al.\ (1989) test was applied to the 
resulting sequence of flares. The result is 
shown as the dashed curve in Fig.~1. Once again the distribution of $H_i$ is
significantly different from that expected from local Poisson behaviour.
The distribution is clearly very similar to that found for the
GOES data. Although this simulation is crude, it shows that 
a Poisson simulation with obscuration produces an effect mimicking 
that reported by Lepreti et al.\ (2001) for the real GOES flares.

\section{Solar cycle variation}

\noindent
Boffeta et al.\ (1999) argued that the observed power-law tail in
the WTD for the GOES events is ``robust,'' and even claimed that the power 
law in waiting times is ``as firmly established as the power laws observed
for total energy, peak luminosity, and time duration.'' Lepreti et al.\ 
(2001) went further by assigning special significance to the 
fitting of the power law by a L\'{e}vy function. The value of the power-law 
index is important to this fit.

A feature of the other power laws (energy, peak flux, and possibly
duration) observed for solar flares is that they 
{\em do not vary substantially with the 11-year solar cycle}
\citep{den85,lu&ham91}, although 
evidence has been presented for small variations \citep{bai93}. 
Here we show that, by contrast, the power law in waiting times is
strongly dependent on the phase of the cycle.

Fig.~2 shows the WTDs for the maximum and minimum phases of the solar
cycle. Maximum phase has been defined as times when the smoothed 
monthly sunspot number\footnote{This data, as well as the GOES catalog,
is available from ${\rm ftp\!\!:\!\!//ftp.ngdc.noaa.gov/STP/SOLAR\_DATA.}$}
is above 125, and minimum has been defined as times when the sunspot
number is below 30. These definitions are arbitrary, but they provide an 
objective choice of cycle phase, i.e.\ they are made without reference to
the flare data. The two histograms are clearly different. The histogram 
extending to larger waiting times corresponds to minimum phase. This is 
not surprising: the mean rate of flaring varies by more than an order of 
magnitude over the cycle, and since the mean rate is the reciprocal of the 
average waiting time, it follows that the WTD must vary with the cycle. 
Fits to the power-law tails ($>10\,{\rm h}$) of the WTDs shown in Fig.~2 
give power-law indices of $-3.2\pm 0.2$ and $-1.4\pm 0.1$ for the maximum 
and minimum phases respectively. The solid curves shown in Fig.~2
correspond to the piecewise constant Poisson 
model~(\ref{eq:varying_poisson}) with rates and times obtained by the
Bayesian procedure \citep{whe00}. This model accounts for the qualitative
shape of the WTDs in each case, although clearly there is some discrepancy,
which is most likely due to the failure of the Bayesian procedure to
account for all of the rate variation.  

\section{Conclusions}

\noindent
Lepreti et al.\ (2001) demonstrated a departure from Poisson
statistics in the GOES catalog of flares, and attributed this to ``time
invariance'' in the flaring history. Here an alternative explanation is 
given, namely that the flux increase due to large flares reduces the
likelihood of detecting subsequent smaller flares. This selection effect
(termed obscuration) introduces a departure from Poisson statistics because 
it effectively means that the rate decreases when a large flare occurs, and 
hence the flares in the GOES catalog are not strictly independent. 
A simulation confirms that
obscuration produces an effect very similar to that reported by Lepreti et
al.\ (2001). This study underlines the need to consider all potential 
biases in the statistical study of flares.  

The power-law tail to the WTD constructed from the GOES catalog has been 
shown to vary with time, and in particular with the solar cycle. On this 
basis it is difficult to argue that the power law is ``robust,'' in marked 
contrast with the other power laws reported in flare statistics.

The picture emerging from this study and others (e.g.\ Wheatland 2001) is 
that flares occur as a time-varying, or non-stationary Poisson process. 
The observed WTD 
depends on the observed rates of flaring \citep{whe00} which
naturally accounts for the variation of the WTD over the solar cycle. 
There is no particular significance to the appearance of a power law in the
WTD. Indeed, adopting the continuous version of~(\ref{eq:varying_poisson}),
viz.\
\begin{equation}\label{eq:lap}
P(\Delta t)=\frac{1}{\lambda_0}\int_{0}^{\infty}f(\lambda)\lambda^2
  e^{-\lambda\Delta t}\,d\lambda,
\end{equation}
where $f(\lambda) d\lambda$ is the fraction of time that the rate is 
in the range $\lambda$ to $\lambda +d\lambda$ and
$\lambda_0=\int_{0}^{\infty}\lambda f(\lambda)\,d\lambda$
is the mean rate of flaring, the asymptotic behaviour of $P(\Delta t)$ 
is obtained by Taylor expanding $f(\lambda )$ about $\lambda=0$. This
gives 
\begin{equation}\label{eq:asympt}
P(\Delta t)=\frac{2f(0)}{\lambda_0}(\Delta t)^{-3}
  +\frac{6f^{\prime}(0)}{\lambda_0}(\Delta t)^{-4}+...
\end{equation}
Equation~(\ref{eq:asympt}) suggests that the asymptotic form of 
$P(\Delta t)$ will always be a power law, with an index of $-3$. In
assessing this result it should be remembered that~(\ref{eq:lap}) is an 
approximate expression (it assumes the rate does not varying greatly 
during a waiting time). Lepreti et al.\ (2001) appeared to forget this point
in criticising the WTD obtained by Wheatland (2000) with 
an exponential form in~(\ref{eq:lap}). 

Interestingly, Earthquake data also often exhibit power-law WTDs 
due to non-stationarity of Earthquake sequences. Utsu (1970) states
that there is ``no primary importance'' to the appearance of a power law,
but the power law exhibited by Earthquake size distributions (the
Gutenberg-Richter relationship) is considered important.
The situation seems to be closely analogous to that for flares.
 
The results of this paper appear to be consistent with an avalanche model 
of flaring subject to a time-varying rate of driving \citep{nor&01}. 
However, the appearance of Poisson statistics does not exclude other models 
with similar general hypotheses concerning energy storage. Indeed, any steady 
state system with rates of downward transition in energy (flares) 
depending only on the size of the transition, and with a mean energy
larger than the majority of flares, will exhibit Poisson statistics
\citep{cra&whe01}.

The author acknowledges the support of an Australian Research Council QEII
Fellowship.

\newpage

\begin{figure}[H]
\plotone{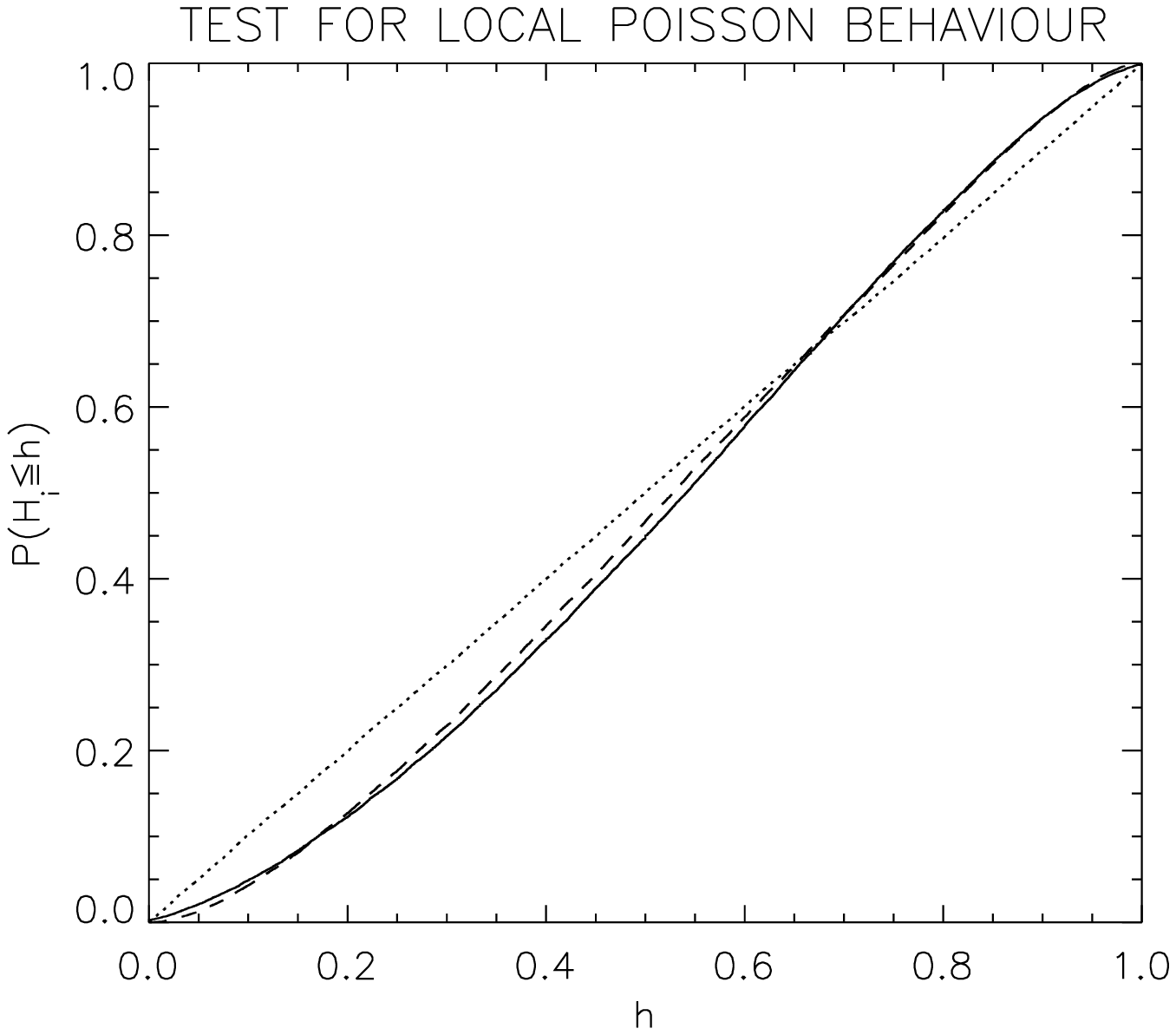}
\caption{The Bi et al.\ (1989) test for Poisson behaviour, 
applied to the GOES data (solid), to a Poisson simulation (dotted), and
to the same Poisson simulation including obscuration (dashed).}
\end{figure}

\begin{figure}[H]
\plotone{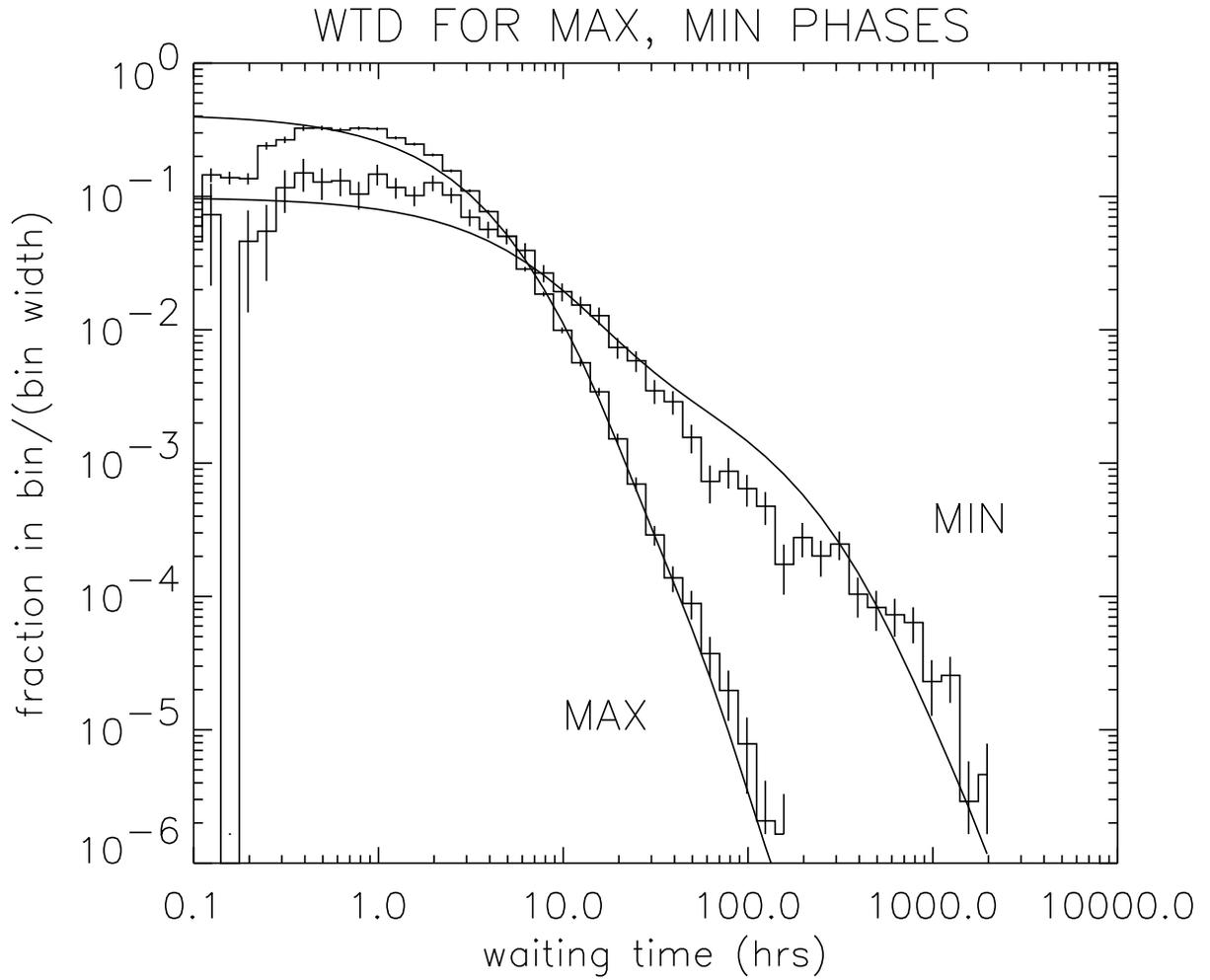}
\figcaption{WTDs for the maximum and minimum phases of the solar 
cycle.}
\end{figure}

\end{document}